\documentclass[aps,twocolumn,floatfix,showpacs,prl]{revtex4-1} 
\usepackage{graphicx}
\usepackage{amsmath}
\usepackage{amsfonts} 
\usepackage{bm}
\usepackage{color}
\usepackage{subfigure}
\usepackage{txfonts}

\begin{document}
\title{Proposal for Coherent Coupling of Majorana Zero Modes and Superconducting Qubits Using the $4\pi$-Josephson Effect}
\author{David Pekker,$^1$ Chang-Yu Hou,$^{1,2}$ Vladimir E. Manucharyan,$^3$ and Eugene Demler$^4$}
\affiliation{
$^1$Department of Physics, California Institute of Technology, Pasadena, California 91125
\\
$^2$Department of Physics and Astronomy, University of California at Riverside, Riverside, California 92521
\\
$^3$Society of Fellows, Harvard University, Cambridge, Massachusetts 02138, USA
\\
$^4$Physics Department, Harvard University, Cambridge, Massachusetts 02138, USA
}
\begin{abstract} We propose to use an ancilla fluxonium qubit to interact with a Majorana qubit hosted by a topological one-dimensional wire.  The coupling is obtained using the Majorana qubit-controlled $4\pi$ Josephson effect to flux bias the fluxonium qubit. We demonstrate how this coupling can be used to sensitively identify the topological superconductivity, to measure the state of the Majorana qubit, to construct 2-qubit operations, and to implement quantum memories with the topological protection. 
\end{abstract}

\pacs{74.20.Mn,73.63.Nm,74.50.+r}
\maketitle

Topological superconducting wires have received considerable experimental and theoretical attention because Majorana zero-energy modes robustly appear at the ends of these wires. These exact zero-energy modes can potentially be used for decoherence-free quantum computation~\cite{Kitaev2001,Nayak2008,Beenakker2013,Alicea2012}. Recent observations of the zero-bias anomaly in proximity-coupled semiconductor-superconductor nanowire devices~\cite{Mourik2012,Das2012,Deng2012} could be interpreted as evidence of Majorana zero-modes~\cite{Law2009,Sau2010b,Pikulin2012}. A more compelling signature of the topological superconductivity is the unusual Josephson current-phase relation. The current-phase relation has two dominant periodicities: a conventional $2\pi$-periodic Josephson current~\cite{Beenakker1991} and an unconventional $4\pi$-periodic component, associated with a pair of Majorana modes near the junction~\cite{Fu2009,Lutchyn2010}.

Although several experiments have already reported indirect evidence for the $4\pi$-Josephson effect in topological junctions~\cite{Williams2012,Rokhinson2012}, the unambiguous detection of the $4\pi$-periodic component may prove difficult. First, realistic topological wires can have many transverse (odd) channels~\cite{Potter2010}. Since each channel contributes to the $2\pi$-supercurrent but only a single channel is topological and contributes to the $4\pi$-supercurrent, the former will typically dominate. This reduces the relative signal strength in proposals related to the phase-biased or voltage-biased junctions~\cite{Kwon2004a,Fu2009}. Second, both coherent and incoherent fluctuations of the parity of the Majorana modes will make the dc-signal $2\pi$-periodic, further complicating the interpretation~\cite{Badiane2011}.

\begin{figure}
\includegraphics[width=8.6cm]{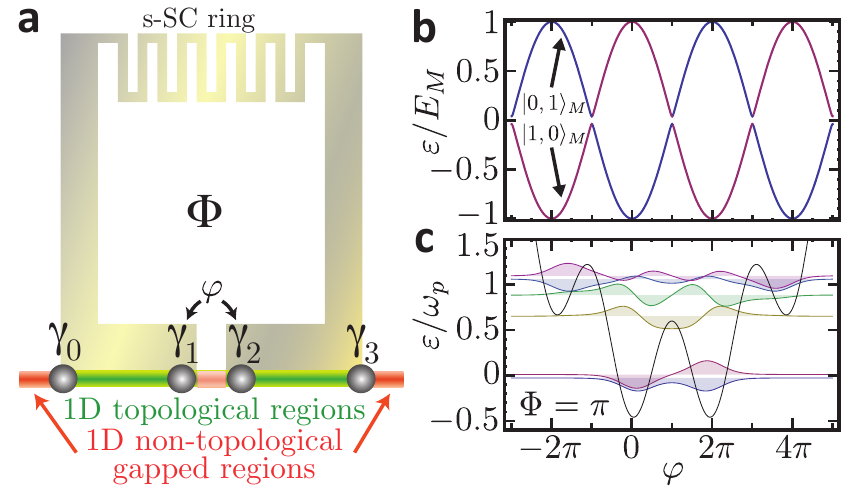}
\caption{(a) Schematic of a Majorana qubit coupled to a fluxonium qubit (see text). (b) Energy spectrum of a Majorana qubit as a function of superconductor phase-difference $\varphi$ with total odd parity. (c) The effective potential energy (black curve) and the lowest six eigenfunctions of a convention fluxonium qubit with $\Phi=\pi$.}
\label{fig:schematic}
\end{figure}

In this Letter, we consider a device made up of a Majorana qubit~\cite{Kitaev2001,Nayak2008} coupled to a fluxonium qubit~\cite{Manucharyan2009} schematically depicted in Fig.~\ref{fig:schematic}. The device consists of a broken superconducting ring coupled to a one-dimensional (1D) quantum wire. Two 1D topological superconducting segments are induced due to the proximity effect. The section of the wire bridging the break of the superconducting ring remains non-topological and acts as a weak link (Josephson junction)  between the two topological regions. Consequently, four Majorana modes, two near the weak link ($\gamma_{1}$, $\gamma_{2}$)~\cite{Fu2009,Lutchyn2010} and two located at the far ends of the topological segments ($\gamma_{0}$, $\gamma_{3}$), appear in the 1D wire and form a topological qubit. The fluxonium qubit requires a large ring inductance $L$ together with a small junction capacitance $C$. In this setting, the phase $\varphi$ across the junction is not pinned by the externally applied flux $\Phi$ through the loop, but instead it can fluctuate quantum-mechanically. The key effect that we exploit here is the direct coupling of the microscopic Majorana modes to the macroscopic flux in the superconducting loop via coherent quantum phase-slips~\cite{Manucharyan2009, Pop2012, Astafiev2012, Manucharyan2012} across the topological Josephson junction.

There have been two types of proposals for hybridizing Majorana and superconducting qubits. In the first type, a conventional superconducting electrometer, such as a top-transmon qubit~\cite{Hassler2011} or any other device based on the Aharonov-Casher effect~\cite{Hassler2010,Bonderson2011}, was suggested. In the second type, the coupling of a pair of Majorana modes, localized inside a pair of trijunctions, was perturbatively tuned by small (compared to $2\pi$) phase variations produced by a nearby flux qubit~\cite{Jiang2011}. Here we describe a third type, in which a pair of Majorana modes is located inside a Josephson junction undergoing quantum phase-slips. Hence, we have to consider the junction in the extreme quantum limit: the phase across the junction is no longer set externally nor does it fluctuate with small (compare to $2\pi$) excursions; instead, the Majorana modes in the junction experience strong phase fluctuations of order $2\pi$. As a result, the microwave spectrum of the hybrid device is external flux-controlled, offset charge insensitive, and strongly dependent on the parity state of the Majorana qubit.

The main results of the paper are summarized as follows. As flipping the parity of the Majorana modes near the junction changes the direction of the $4\pi$-supercurrent, it effectively flux biases the superconducting ring by $2 \pi$ and alters the fluxonium qubit spectrum. We point out how to take advantage of this effect (a) to detect topological wires even with a very small $4\pi$-periodic component and (b) to read the state of the Majorana qubit. Since reversing the direction of the $4\pi$-periodic Majorana supercurrent is equivalent to changing $\varphi$ by $2\pi$, phase slips will hybridize Majorana and fluxon modes. This hybridization becomes the crucial ingredient for (c) implementing ``controlled-NOT'' (CNOT) operations between the Majorana states and fluxon states by simple microwave pulses.

The Hamiltonian $H_{M-F}$ describing the interaction between Majorana and fluxonium qubits can be split into two terms:
\begin{equation}
H_{M-F} = H_M(\varphi) + H_F(\varphi, \Phi),
\label{eq:M+F}
\end{equation}
where $H_M$ describes the Majorana qubit and the $4\pi$-Josephson effect and $H_F$ governs the macroscopic quantum dynamics of the fluxonium qubit.

\textit{Majorana parity qubit} -- \/
In the absence of the $H_F$ term, $\varphi$ can be treated as a parameter. A generic phenomenological model for the coupled Majorana modes is given by~\cite{Kitaev2001}
\begin{equation}
\label{eq:H_M}
H_M=g^{\ }_{01} i \gamma_0 \gamma_1 + E^{\ }_M i \gamma_1 \gamma_2 \cos(\varphi/2+\Theta^{\ }_M)
 + g^{\ }_{23} i \gamma_2 \gamma_3.
\end{equation}
Here, $g_{ij}$ is the coupling between the Majorana modes $\gamma_i$ and $\gamma_j$, and $E_M$ and $\Theta_M$ are the strength and a phase shift of the $4\pi$-Josephson effect. Typically, $g_{ij}\ll E_M$ as the Majorana mode coupling decays exponentially with respect to the bulk gap of the wire. $g_{03}$ is negligible when $\gamma_0$ and $\gamma_3$ are far apart.

The physical origin of the $4\pi$-periodic Josephson effect comes from the boundary conditions for Bogoliubov quasiparticles. While a shift of $\varphi$ by $2\pi$ must leave the boundary conditions for the superconducting order parameter invariant, the quasiparticles see only ``half" of this phase, and are therefore invariant only to shifts of $\varphi$ by $4\pi$. The $4\pi$-Josephson effect is a consequence of the coupling of $\gamma_1$ and $\gamma_2$ via the junction, thus its strength $E_M$ is related to the transparency of the junction and does not scale with the number of transverse channels.

The phase shift $\Theta_M$ can be finite since the wave functions of the operators $\gamma_1$ and $\gamma_2$ are generically unrelated. To have $\Theta_M=0$, the phases of wave functions need to be fixed independently. This can be accomplished, for instance, by ensuring that the Hamiltonian for the topological wire segments is real~\cite{Tewari2012} (see supplemental material). We will set $\Theta_M=0$ except in the discussion of the two-qubit operations where a finite $\Theta_M$ becomes a useful resource.

In terms of conventional (complex) fermions, $H_{M}$ is
\begin{align}
\label{Eq:H_Mc}
 H_{M,c}=E_M \Big(c_w^\dagger c_w - \frac{1}{2}\Big)\cos \frac{\varphi}{2}- \lambda_+ c_w^\dagger c_e -\lambda_- c_w^\dagger c_e^\dagger + \text{h.c.},
\end{align}
where $\lambda_\pm=2(g_{01}\pm g_{23})$; $c_w=\left(\gamma_1 + i \gamma_2\right)/2$ and $c_e=\left(\gamma_3 + i\gamma_0\right)/2$ describe a local fermion at the weak-link ($w$) and a  ``split" fermion at the outer ends of the wires ($e$), respectively. The Hilbert space of $H_{M,c}$ can be defined by the fermion occupation numbers of $n_w = c_{w}^{\dagger}c_{w}$ and $n_e = c_{e}^{\dagger}c_{e}$ as $|n_w, n_e\rangle_M$ with $n_w, n_e = \{0,1\}$.

As $H_{M,c}$ conserves the \textit{combined} fermion parity $n_w+n_e$, the states [$|0, 1\rangle_M$, $|1, 0\rangle_M$] and [$|0, 0\rangle_M$, $|1, 1\rangle_M$] form two decoupled (odd and even) sectors. Therefore, basis states of the Majorana parity qubit can be defined as $|n_w =0\rangle$ and $|n_w = 1\rangle$ that correspond to the two parities of $n_w$ with a fixed combined parity $n_w + n_e$. The two-level spectrum of the odd sector is plotted in Fig.~\ref{fig:schematic}b (the even sector being exactly the same). Due to the couplings $g_{ij}$, the states $|0,1\rangle_M$ and $|1,0\rangle_M$ anticross at $\varphi = \pi$ by $\lambda=\lambda_+$ [$\lambda=\lambda_-$ for the even sector]. Consequently a slow passage through the anticrossing coherently flips both $n_w$ and $n_e$.

\textit{Fluxonium qubit -- \/}
$H_F$ in Eq.~(\ref{eq:M+F}) turns $\varphi$ into a quantum-mechanical variable
\begin{equation}
H_F(\varphi, \Phi) =- 4E_C \, \partial_\varphi^2 + \frac{1}{2} E_L (\varphi-\Phi)^2-E_J \cos\varphi ,
\label{eq:Fluxonium}
\end{equation}
where $E_J$ is the Josephson energy, $E_C = e^2/2C$ is the charging energy, $E_L = (\Phi_0/2\pi)^2/L$ is the inductive energy, and $\Phi$ is measured in units of $\Phi_0/2\pi=\hbar/2e$. $H_F$ is formally equivalent to the Hamiltonian of a particle with a coordinate $\varphi$ and a mass proportional to $C$, traveling in an effective potential (Fig.~\ref{fig:schematic}c) defined by $E_L$, $E_J$, and $\Phi$. The inductance $L$ must be sufficiently large, such that $E_J > E_L$, to ensure a set of well-defined local potential minima spaced approximately by $2\pi$. 

Classically, i.e., for $C \rightarrow \infty$, the phase $\varphi$ localizes in one of the Josephson wells and vibrates at the plasma frequency $\omega_p \approx \sqrt{8 E_J E_C}$. The presence of quantum tunneling (finite $C$) allows $2\pi$ phase-slips between the adjacent wells. At the maximal frustration of $\Phi =  \pi$, as shown in Fig.~\ref{fig:schematic}c, the two lowest eigenstates of $H_F$ correspond to equal superpositions of the states with $\varphi \approx \{0,\; 2\pi\}$. Coherent oscillations between such states correspond to a flux quantum -- ``fluxon'' -- entering and leaving the loop (charging or discharging the inductance). The fluxon picture make sense only when the $2\pi$-slip events are relatively rare, which requires $\sqrt{8E_J/E_C} \sim 1$~\cite{Matveev2002} (see Ref.~\onlinecite{Pekker2012} for $\sqrt{8E_J/E_C} \gg 1$ regime). Consequently, because of the large $L$, $\varphi$ fluctuates with typical deviations comparable to $2\pi$.

\begin{figure*}
\includegraphics[width=17.6cm]{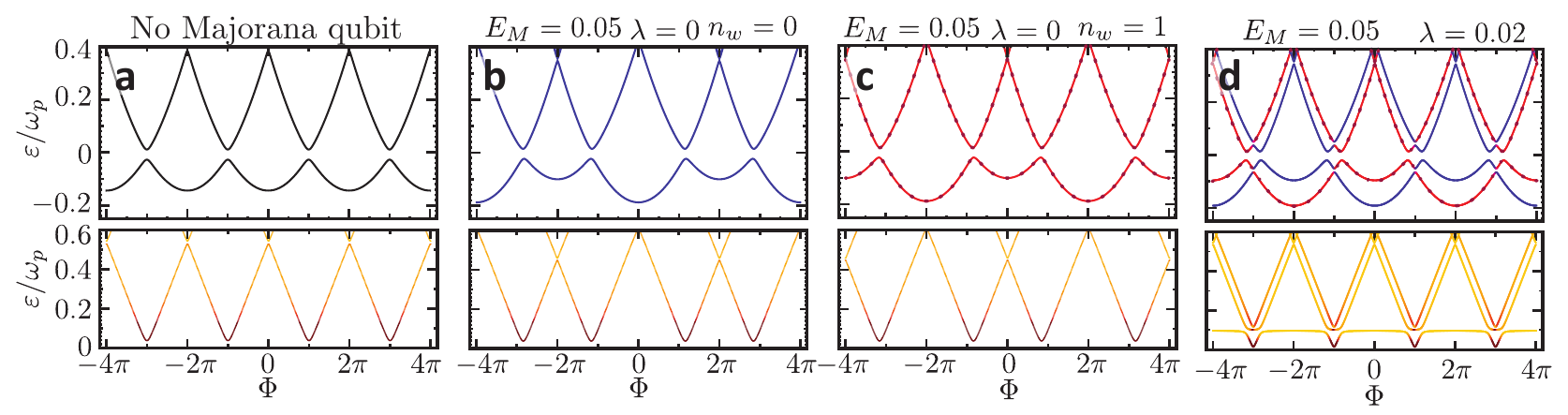}
\caption{Upper panels: Spectra of Majorana-fluxonium device, Eq.~\eqref{eq:FluxoniumMajorana}, are plotted as a function of $\Phi$ for four scenarios [in the presence of Majorana qubit, the blue (red-dotted) colored curve indicates $\langle n_w \rangle=0$ ($\langle n_w \rangle=1$)]. Lower panels: Corresponding transition frequencies between the ground state and the first few excited states [darker color indicates larger transition rate]. (a) pure fluxonium qubit case; (b) and (c) topological case with $E_M/\omega_p=0.05$, $\lambda=0$ and fixed $n_w$: $n_w=0$ and $1$, respectively; (d) topological case with parity fluctuation $\lambda/\omega_p=0.02$. (In all cases: $E_J/\omega_p=0.6$, $E_L/\omega_p=0.03$ and $\Theta_M=0$)
}
\label{fig:flow}
\end{figure*}

\textit{Coupling Majoranas to fluxons -- \/} 
The strong quantum fluctuations of $\varphi$ at $\Phi \approx \pi$ change qualitatively the Majorana qubit spectrum. Combining Eqs.~(\ref{Eq:H_Mc}, \ref{eq:Fluxonium}), we get 
\begin{equation}
H_{M-F} = H_F (\varphi, \Phi) \openone - E_M \cos (\varphi/2) \sigma_z + \lambda \sigma_x.
\label{eq:FluxoniumMajorana}
\end{equation}
where $\openone$ and $\sigma_{\{x,y,z\}}$ are the $2\times2$ identity and Pauli matrices acting on the $n_w=\{0,1\}$ basis. $H_{M-F}$ can be diagonalized numerically (Fig.~\ref{fig:flow}). We will focus on the practically important case of $E_M < E_J$. As fluxonium requires $E_L < E_J$, we can always select the inductance, $L$, such that $E_M < \pi^2 E_L < E_J$. In that case, an effective Hamiltonian for the low-energy spectrum of $H_{M-F}$ reads
\begin{align}
\label{eq:FluxonMajorana-eff}
H^{\rm{eff}}_{M-F} = \frac{E_L(2\pi n_\varphi - \Phi)^2}{2}  - \frac{E_S}{2} \sum_{a=\pm} T^{a}_{n_{\varphi}} + 4 E_M (-1)^{n_\varphi}\sigma_z
 +  \lambda \sigma_x, \nonumber
\end{align}
where $n_\varphi$ is the fluxon number operator, $T^{\pm}_{n_{\varphi}}|n_\varphi\rangle = |n_\varphi \pm1\rangle$, and $E_S = E_S(E_J, E_L, E_M)$ is the $2\pi$ phase-slip amplitude~\cite{Mooij2006,Koch2009}. The $E_M$-term couples fluxon states $|n_\varphi\rangle$ to the Majorana qubit states $|n_w\rangle$ which we now describe in the combined basis $|n_\varphi, n_w\rangle$.

At $E_M = 0$, we recover the $2\pi$-periodic spectrum of the fluxonium qubit, which consists of the fluxon parabolas spaced in $\Phi$ by $2\pi$ and anticrossed at $\Phi = \pm \pi, ...$ (Fig.~\ref{fig:flow}a). For $E_M \neq 0$ and $\lambda =0$, there are two sets of fluxon states $|n_\varphi, 0\rangle$ and $|n_\varphi, 1\rangle$ (Fig.~\ref{fig:flow}b,c). Within each set, there is a $2 E_M$ offset between the fluxon parabolas with $n_\varphi$ even[odd] and odd[even] for $n_w = 0$[$n_w = 1$]. Consequently, the fluxon anticrossings now occur away from $\Phi = \pm \pi$, rendering all transition energies to be $4\pi$-periodic. The condition $E_M < \pi^2 E_L$ ensures that the anticrossing of the states $n_\varphi$ and $n_\varphi +1$ occurs at a lower energy than the crossing between the states $n_\varphi$ and $n_\varphi+2$, which allows us to neglect direct $4\pi$ phase-slips.

We observe that the ground state of $H^{\text{eff}}_{M-F}$ has a degeneracy at $\Phi = \pi$ corresponding to the crossing of the states $|n_\varphi = 0, n_w = 0\rangle$ and $|n_\varphi = 1, n_w = 1 \rangle$. This is evident from superimposing the spectra in Fig.~\ref{fig:flow}b and Fig.~\ref{fig:flow}c. The crossing is robust to parameter variations as long as the fermion parity $n_w$ is fixed, similarly to the crossing in the spectrum of the conventional Majorana qubit at $\varphi = \pi$. The doubly degenerate ground states can be split by a process flipping the fermion parity in the junction region simultaneously with changing the fluxon number in the loop by a unity, which requires $E_S \neq 0$ and $\lambda \neq 0$. Then, in the vicinity of $\Phi=\pi$, fluxon fully hybridizes with the Majorana fermion parity, making a ``fluxpariton.'' Thus, the symmetric and antisymmetric superpositions of wave functions $|n_\varphi = 0, n_w = 0\rangle$ and $|n_\varphi = 1, n_w = 1 \rangle$ become the new ground and first excited states. The splitting $2g_{M-F}$ between these states is $2 g_{M-F} \approx \lambda E_S/\sqrt{4 E_M^2 + E^2_S}$ and restores $2\pi$-periodic spectrum as shown in Fig.~\ref{fig:flow}d.

{\it Controlling the Majorana qubit with fluxonium -- \/} can be performed using microwave spectroscopy of the type used in conventional fluxonium qubits~\cite{Manucharyan2012}. Before discussing coherent oscillations between the Majorana and fluxonium qubits, we shall comment on incoherent processes. In practice, the combined fermion parity $n_w+n_e$ can fluctuate incoherently at some time scale $t_{qp}$ due to out-of-equilibrium processes~\cite{Aumentado2004}, known as quasiparticle poisoning. Therefore, to make the spectroscopy possible, $t_{qp}$ must be longer than the transition time $t_{\pi}$ (the time to generate a $\pi$-pulse). We note that $t_{\pi}$ can be tuned over a wide range, as it is proportional to the driving power and the transition matrix element of $\varphi$ (see Fig.~\ref{fig:flow}). In a typical superconducting qubit $t_{\pi}$ is about $1\sim10$ ns, while $t_{qp}$ is in the range of $10~\mu\text{s}\sim 1~\text{ms}$~\cite{Manucharyan2012, Riste2013}.

How can we address the fundamental experimental issue of identifying Majorana modes? The presence of a finite $E_M$-term results in the appearance of two distinct features even if the fully coupled Majorana-fluxon spectrum is $2\pi$-periodic. From the lower panel of Fig.~\ref{fig:flow}d, we observe (1) a nearly $\Phi$-independent transition at $2 E_M$, which anticrosses with the fluxonium ``zigzag''-shaped line around $\Phi = \pi$, and (2) the splitting of the zigzag line by $2 E_M$. Remarkably, $E_M$ as small as the line width of the fluxonium qubit can be resolved. 

When the $\lambda$ coupling becomes negligible, we expect to observe superimposed spectra of Fig.~\ref{fig:flow}b or Fig.~\ref{fig:flow}c due to the fluctuation of the $n_w$ occupation. With a fixed value of $\Phi$, the incoherent fluctuation of the fermion parity alters the resonance frequency, depending on the occupation of $n_w$. Hence, by monitoring the switching time of the resonance frequency, we can infer the quasiparticle poisoning time $t_{qp}$.

The proposed device allows us to prepare an arbitrary state of the Majorana qubit by (i) initializing the coupled Majorana-fluxonium system in the ground state using standard superconducting qubit techniques~\cite{Geerlings2013}, and (ii) applying microwave pulses at the frequency of the transition $|0, 0\rangle \rightarrow |0, 1\rangle$. To read out the state of the Majorana qubit, one can use  spectroscopy (away from anticrossings) to project onto a definite $n_w$ parity state (blue and red-dotted lines in Fig.~\ref{fig:flow}d).

\begin{figure}
\includegraphics[width=8.5cm]{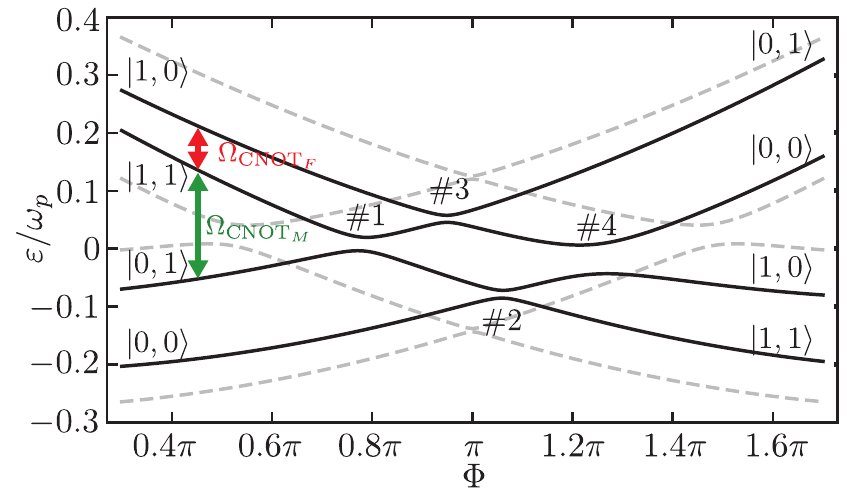}
\caption{Device spectrum as a function of $\Phi$ zoomed in on the four lowest eigenvalues. $\Theta_M=0$ (dashed lines) is compared to $\Theta_M=-0.35 \pi$ (solid lines). The eigenvalues are labeled using the uncoupled qubit eigenbasis $|n_{\varphi}, n_{w} \rangle$. Level crossings are numbered $\#1$ to $\#4$, and driving frequencies needed to implement $\text{CNOT}_\text{F}$ and $\text{CNOT}_\text{M}$ are labeled. ($E_J=0.6\omega_p$, $E_L/\omega_p=0.03$, $E_M/\omega_p=0.15$, $\lambda/\omega_p=0.02$)}
\label{fig:spectrumB}
\end{figure}

{\it Two-qubit operations -- \/} like the two-qubit controlled gates are essential for quantum computing. In our device, the CNOT gates can be implemented by a single $\pi$-pulse in the presence of a finite phase-shift $\Theta_M$ [$E_M \cos (\varphi/2) \rightarrow E_M \cos (\varphi/2 + \Theta_M)$ in Eq.~\eqref{eq:FluxoniumMajorana}]. We plot the four lowest eigenvalues comprising the Hilbert space of the two qubits in Fig.~\ref{fig:spectrumB} as a function of $\Phi$. The level crossings \#1 and \#4 are due to pure $2\pi$-phase slips while the crossings \#2 and \#3 are due to Majorana-fluxon hybridization. $\Theta_M\neq0$ offsets the crossings \#2 and \#3 away from $\Phi=\pi$, see Fig.~\ref{fig:flow}d, and lifts the near degeneracies of the transition frequencies $\Omega_{|0,0\rangle \leftrightarrow |0,1\rangle }$ and $\Omega_{|1,1\rangle \leftrightarrow |1,0\rangle }$ thus allowing them to be addressed independently.

The Majorana qubit controlled CNOT$_{\text M}$ gate corresponds to a $\pi$-pulse with frequency $\Omega_{|0,1\rangle \leftrightarrow |1,1\rangle}$ as depicted by the green arrow in Fig.~\ref{fig:spectrumB}. Similarly, the fluxonium qubit controlled CNOT$_{\text F}$ gate can be implemented by a $\pi$-pulse with frequency $\Omega_{|1,1\rangle \leftrightarrow |1,0\rangle}$ (red arrow in Fig.~\ref{fig:spectrumB}). A swap gate, which can be implemented by performing CNOT$_{\text F}$, CNOT$_{\text M}$ and CNOT$_{\text F}$ in sequence~\cite{Nielsen2000}, can be used to move quantum information into and out of the topologically protected Majorana qubit and hence to implement a partially topologically protected quantum memory.

{\it Experimental requirements -- \/} to couple the Majorana states to fluxons are fully compatible with the two widely discussed strategies for the implementation of a 1D topological superconductor: semiconducting nanowires with strong spin orbit scattering~\cite{Lutchyn2010,Oreg2010,Alicea2011} and quantum-spin-Hall effect edge states~\cite{Fu2009,Nilsson2008}. In both cases, the broken superconducting ring depicted in Fig.~\ref{fig:schematic}a is made from an s-wave superconductor and serves two functions: it induces the gap in the wire/edge by proximity effect, and provides an inductance $L$ for the fluxons. $\Theta_M$ can be tuned using a magnetic field (see supplemental material), and $C$ can be tuned using an external electromagnetic structure~\cite{Koch2007}.

Typical parameters of the fluxonium qubit are such that $E_L/h \sim (0.1 - 1) \rm{GHz}$, $E_J/h \sim (5 - 50) \rm{GHz}$, and $E_C/h \sim {0.5 - 5} \rm{GHz}$. To provide a large $L$, a good choice for the superconductor can be NbN, which has high kinetic inductance and $T_c\sim10 \rm{K}$~\cite{Annunziata2010}. Both $E_J$ and $E_M$ depend on the transparency of the topological junction, which can be tuned by the local gating of the junction region. Most importantly, all electronic gaps need to be sufficiently large $\gg \omega_p$ to suppress exciting quasiparticles during operations~\cite{Takei2013}.

{\it Concluding remarks -- \/} We showed that the spectrum of the superconducting fluxonium qubit is highly sensitivity to the presence of the Majorana modes in the Josephson junction. To compare the sensitivity of our device to an experiment that would directly measure the $4\pi$ signal in the current-phase relation of a topological Josephson junction, we remark that the fluxonium transition shifts by $1~\rm{MHz}$ per $100~\rm{fA}$ of $4\pi$ Josephson current. As the transition frequency resolution of our device is limited only by the fluxonium quality factor, we can expect a sub-$100~\rm{fA}$ sensitivity to $4\pi$ supercurrents~\cite{Manucharyan2012}. 
The key effect responsible for such high sensitivity is the coupling of Majorana modes with fluxons in the large-inductance fluxonium loop. This coupling can be used to hybridize the two qubits and perform non-topological quantum manipulations of the Majorana qubit. Further, large-inductance loops can be readily incorporated into the general scheme of gate-controlled nanowire networks to compliment braiding operations with the still required non-topological operations: state initialization, readout, and single-qubit rotations. Finally, the high sensitivity of the proposed device can be used to unambiguously identify topological superconducting wires.

{\it Acknowledgements -- \/} It is our pleasure to thank J. Sau, D. Abanin, C. Marcus and A. Akhmerov for useful discussions. We would like to thank the KITP for its hospitality. DP thanks the Lee A. DuBridge fellowship and the IQIM, CYH thanks DARPA-QuEST program, EAD thanks the Harvard-MIT CUA, NSF Grant No. DMR-07-05472,  ARO-MURI on Atomtronics, and ARO MURI Quism program.

\bibliographystyle{apsrev}
\bibliography{fluxonium}

\newpage

\section{Supplemental Material}

\subsection{Diagonalizing the effective Hamiltonian}
In this supplement we provide explicit details on how we diagonalize the effective Hamiltonian Eq.~(5). We begin by rewriting the effective Hamiltonian in terms the harmonic oscillator frequency $\omega=\sqrt{ 8 E_L E_C}/\hbar$ and length $s=\sqrt{\frac{8 E_c}{\hbar \omega}}$ scales
\begin{eqnarray}
\label{eq:modelBHs}
h_{\text{eff}}&=&\left[ -\frac{1}{2} \partial_\phi^2+\frac{1}{2} \phi^2-e_J \cos(s\phi+\Phi)\right] \openone  
\\
&&\quad\quad - e_M \cos \left( \frac{s\phi}{2}+\frac{\Phi}{2} + \Theta_M \right) \sigma_z + \ell \sigma_x. \nonumber
\end{eqnarray} 
where, $h_\text{eff}=H_\text{eff}/\hbar \omega$, $\phi=(\varphi-\Phi)/s$, $e_J=E_J/\hbar \omega$, $e_M=E_M/\hbar \omega$, and $\ell=\lambda/\hbar \omega$.

In order to diagonalize the effective Hamiltonian, Eq.~\eqref{eq:modelBHs}, we find it useful to use the harmonic oscillator states, corresponding to the first two terms of the Hamiltonian, as our basis. Explicitly, these states are
\begin{align}
\chi_n(\phi)=\frac{1}{\sqrt{2^n n!}} \left(\frac{1}{\pi}\right)^{1/4} e^{-\phi^2/2} H_n(\phi),
\end{align}
where $H_n(\cdot)$ are the Hermite polynomials. To construct a basis for Eq.~\eqref{eq:modelBHs} we tensor the Harmonic oscillator states with the topological qubit states. In this basis, the effective Hamiltonian Eq.~\eqref{eq:modelBHs} becomes
\begin{align}
h_\text{eff}=\left(
\begin{array}{cc}
h_{+e_M} & \ell \, {\bf 1} \\
\ell \, {\bf 1} & h_{-e_M}
\end{array}
\right),
\end{align}
where ${\bf 1}$ is the identity matrix and
\begin{widetext}
\begin{align}
(h_{\pm e_M})_{ij}&=\left(i+\frac{1}{2}\right) \delta_{i,j} - e_J \left[\cos(\Phi) C_{i,j}(s) - \sin(\Phi) S_{i,j}(s)\right] \mp e_M \left[\cos\left(\frac{\Phi}{2}+\Theta\right) C_{i,j}\left(\frac{s}{2}\right)- \sin\left(\frac{\Phi}{2}+\Theta_M\right) S_{i,j}\left(\frac{s}{2}\right)\right].\nonumber
\end{align}
\end{widetext}
The matrix elements $C_{i,j}(\cdot)$ and $S_{i,j}(\cdot)$ are obtained using formula 7.388 (6,7) of Ref.~\onlinecite{Gradshteyn}
\begin{align}
C_{n,n+2m}(b)&=\langle \chi_n(\phi) | \cos(b \phi) | \chi_{n+2m}(\phi) \rangle \nonumber \\
&\!\!\!\!\!\!\!\!\!\!\!\!= \frac{(-1)^m b^{2m} e^{-\frac{b^2}{4}}\sqrt{2^n n!} L_n^{2m}\left(\frac{b^2}{2}\right)}{\sqrt{2^{n+2m}(n+2m)!}},\\
S_{n,n+2m+1}(b)&=\langle \chi_n(\phi) | \sin(b \phi) | \chi_{n+2m+1}(\phi) \rangle \nonumber \\
&\!\!\!\!\!\!\!\!\!\!\!\!= \frac{(-1)^m b^{2m+1} e^{-\frac{b^2}{4}}\sqrt{2^n n!} L_n^{2m+1}\left(\frac{b^2}{2}\right)}{\sqrt{2^{n+2m+1}(n+2m+1)!}},
\end{align}
where $C_{n,n+2m+1}=S_{n,n+2m}=0$, and we use the property that $C_{i,j}(b)=C_{j,i}(b)$ and $S_{i,j}(b)=S_{j,i}(b)$ for matrix elements with $i>j$. To obtain the lowest $2k$ eigenvalues of the Hamiltonian, we need to keep $2p$ basis elements with $2p \geq 2k$. In practice, we start with $2p=2k$ and increase $2p$ until the first $2k$ eigenvalues have converged (to construct the figures, we have used $k=10$ and $p=30$).

\subsection{Microwave drive}
The transition matrix element can be evaluated by noting that in the un-rescaled basis magnetic field fluctuations enter into the $E_L$ term via
\begin{align}
\frac{1}{2}E_L \left[\varphi-\Phi-\delta_\Phi \cos(\Omega t)\right]^2 \, {\bf 1}.
\end{align}
Expanding this term in small $\delta_\Phi$, we obtain
\begin{align}
\frac{1}{2}\left[E_L \left(\varphi-\Phi\right)^2 - 2 E_L \left(\varphi-\Phi\right) \cos(\Omega t) \delta_\Phi + O(\delta_\Phi^2)\right] {\bf 1}.
\end{align}
Upon rescaling, the linear in $\delta_\Phi$ term (the dipole matrix element) becomes
\begin{align}
\frac{E_L}{\hbar \omega} s \phi  \cos(\Omega t/\omega) \delta_\Phi \, {\bf 1},
\end{align}
which we can evaluate with the help of the relations
\begin{align}
L_{n,n+1}&=\langle \chi_n(\phi) | \phi | \chi_{n+1}(\phi) \rangle = \sqrt{(n+1)/2}\\
\langle \psi_1 | \varphi | \psi_2 \rangle&=\sum_{i=0}^{p-2} \sqrt{\frac{i+1}{2}} \left(\psi_{1,i}^* \psi_{2,i+1} + \psi_{1,i+1}^* \psi_{2,i}\right).
\end{align}

\subsection{Symmetries of topological superconducting wires and the $4\pi$ Josephson effect}
Consider a large topological superconducting ring with a break. Further, suppose that the superconducting order parameter on the two sides of the break differs in phase by $\varphi$. The broken ring hosts two Majorana fermion: $\gamma_1$ to the left of the break and $\gamma_2$ to the right of the break. Now consider bridging the gap with a weak link, which we model using the tunnel Hamiltonian 
\begin{align}
H=- t \sum_\sigma e^{i \varphi/2} c_{1\sigma}^\dagger c_{2\sigma} + \text{h.c.}
\end{align}
where $t$ is the matrix element associated with the weak link, $c^\dagger_{1\sigma}$ [$c_{2\sigma}$] is the electron creation [annihilation] operator with spin $\sigma$ to the left [right] of the break, and we have absorbed the phase difference $\phi$ into a gauge field associated with the weak link. The Majorana fermion operators are related to the electron operators via the coherence factors 
\begin{align}
c^\dagger_{i\sigma}=u_{i\sigma} \gamma_i/2 + \dots \\
c_{i\sigma}=u^*_{i\sigma} \gamma_i/2 + \dots 
\end{align}
where $i=\{1,2\}$, $\dots$ represent the remaining Bogoliubov operators, and we have used the fact that $\gamma_i$ is self-adjoint.
Using the coherence factors $u_{i,\sigma}$ we express the tunneling Hamiltonian 
\begin{align}
H_\text{t}=\frac{ i t \gamma_1 \gamma_2}{2} \left[A_\uparrow \cos\left[
\frac{\varphi}{2}+\chi_\uparrow \right]\!+\!A_\downarrow \cos\left[
\frac{\varphi}{2}+\chi_\downarrow\right]\right] \label{eq:HT2}
\end{align}
where $t$ is the tunnel matrix element across the weak link, $A_\sigma=|u_{1\sigma}||u_{2\sigma}|$, $\chi_\sigma=\arg\left[u_{1\sigma} u^*_{2\sigma}\right]-\pi/2$. As $|u_{1\sigma}|$ is generically unrelated to $|u_{2\sigma}|$ in the presence of disorder, ensuring that $\Theta_M=0$ requires fixing the phases of $u_{i\sigma}$'s independently.

When the Hamiltonian describing the nanowire is completely real (or equivalently completely imaginary), all wave function can be expressed as a real vectors. In particular, the reality condition implies that the Bogoliubov operator $c_w=\gamma_1 + i \gamma_2$ can be described by real coherence factors only. Following the above definitions we obtain $\arg\left[u_{1\uparrow}\right]=\arg\left[u_{1\downarrow}\right]=0$ and $\arg\left[u_{2\uparrow}\right]=\arg\left[u_{2\downarrow}\right]=\pi/2$. Hence $\Theta_M=\arg\left[u_{1\uparrow}\right]-\left[u_{2\uparrow}\right]-\pi/2=0$ and 
\begin{align}
H_\text{t} \approx i t \gamma_1 \gamma_2 \cos(\phi/2) = \left(c^\dagger_w c_w - c_w c^\dagger_w\right) \cos(\phi/2).
\end{align}

{\it Nanowire implementation of a 1D topological superconductor -- \/} the Hamiltonian for the topological superconductor can be written in the form
\begin{align}
H_\text{NW}&=(-\partial_x-\mu) \tau^z -i \alpha \partial_x \sigma^y \tau^z \nonumber \\
&+ B^x \sigma^x \tau^z + B^y \sigma^y + B^z \sigma^z \tau^z - \Delta \sigma^y \tau^y, 
\end{align}
where we use the notation $(u_{i\uparrow}, u_{i\downarrow}, v_{i\uparrow}, v_{i\downarrow})$ for the four component particle-hole spinor, the Pauli matrices $\sigma^{\{x,y,z\}}$ and $\tau^{\{x,y,z\}}$ act on the spin and particle-hole spaces, respectively; $\mu$ is the chemical potential, $\alpha$ is the spin-orbit velocity, $(B^x, B^y, B^z)$ is the Zeeman field vector, and $\Delta$ is the proximity pairing field. We observe that the complex conjugation operator ${\cal K}$ commutes with all but the $B^y$ term in $H_\text{NW}$, and therefore $\Theta_M=0$ in the absence of $B^y$. On the other hand, applying a $B^y$ field can be used to control $\Theta_M$; in the short junction limit the we find $\Theta_M \sim B^y/\Delta$.

{\it Quantum spin-Hall effect edge state implementation of a 1D topological superconductor -- \/} the Hamiltonian for the topological superconductor is generically
\begin{align}
H_\text{QSH-E}=v (i \partial_x) \sigma^z  - \mu \tau^z + \Delta'' \sigma^y \tau^x + B^y \sigma^y + B^z \sigma^z \tau^z.
\label{eq:QSH-E}
\end{align}
We remark that $B^x \sigma^x \tau^z$ and $\Delta' \sigma^y \tau^y$ terms can be obtained using the rotations generated by $\sigma^z \tau^z$ and $\tau^z$, respectively.  We observe that in the absence of  $\mu$ and $B^z$ the Hamiltonian $H_\text{QSH-E}$ anti-commutes with ${\cal K}$ and is therefore completely imaginary. 
At this point, we note that the Hamiltonian Eq.~\eqref{eq:QSH-E} has an additional symmetry: with the exception of the $B^z$ term all terms of $H_\text{QSH-E}$ commute with the operator ${\cal K} \sigma^x \tau^z$. This additional symmetry implies that in the absence of the $B^z$ term $u_{i,\uparrow}=u_{i,\downarrow}^*$ and therefore $\Theta_M=0$ using Eq.~\eqref{eq:HT2}. Taking into account the ${\cal K} \sigma^x \tau^z$ symmetry, we conclude that a magnetic field $B^z$ aligned with the (1D) spin-orbit axis, but not a chemical potential shift $\mu$, will result in $\Theta_M \neq 0$. 

We remark that for topological superconductors with particle-hole symmetry, the reality condition is related to chiral symmetry, see Ref.~\onlinecite{Tewari2012}.

\end{document}